\documentclass{pic2012}
\def\runauthor{R.\ Waldi}
\def\shorttitle{B Decays and CP Violation}
{
\catcode`\@=11
\newif\ifhistopolyline
\newdimen\histowidth
\newdimen\totalhistowidth
\newdimen\histoheight
\dimendef\dimen@@=3
\dimendef\dimen@@i=4
\dimendef\dimen@v=5
\dimendef\dimen@vi=6
\newcount\histoxmin
\newcount\histoxdif
\newcount\histoymin
\newcount\histoydif
\newcount\count@@
\newdimen\histox
\newdimen\histoy
\newdimen\histoxx
\newdimen\histoyy
\newdimen\histoxu
\newdimen\histoyu
\newdimen\histobarwidth
\newdimen\histotickwidth
\newdimen\historulewidth
\expandafter\ifx\csname fiverm\endcsname\relax\font\fiverm = cmr5\fi
\expandafter\ifx\csname sevenrm\endcsname\relax\font\sevenrm=cmr7\fi
\expandafter\ifx\csname tenrm\endcsname\relax\font\tenrm=cmr10\fi
\expandafter\ifx\csname bigrm\endcsname\relax\font\bigrm=cmr10 scaled 1440\fi
\expandafter\ifx\csname linefont\endcsname\relax%
\font\linefont = line10 at 10truept \fi
\newdimen\plotpointradius
\newdimen\plotdotspacing
\newbox\plotpointbox
\histowidth=300\p@
\histoheight=300\p@
\historulewidth=0.8\p@
\histobarwidth=0.4\p@
\histotickwidth=0.4\p@
\def\plotsymbol{\bullet}
\def\histosize#1#2{\histowidth=#1\relax\histoheight=#2\relax}

\def\starthistogram{%
    \hskip 0pt plus 0.5fil\lower\historulewidth\hbox to \historulewidth{%
    \vbox{\offinterlineskip
    \hsize\histowidth
    \advance\hsize\historulewidth
    \advance\hsize\historulewidth
    \hrule height \historulewidth
    \hbox to \hsize{\vrule height \histoheight width \historulewidth
     \hfil\vrule width \historulewidth}
    \hrule height \historulewidth}\hss}%
    \historescale}
\def\historescale#1#2#3#4{%
    \histoxmin=#1\relax
    \histoxdif=#2\advance\histoxdif by -\histoxmin
    \histox=\histowidth
    \divide\histox by \histoxdif
    \histoymin=#3\relax
    \histoydif=#4\advance\histoydif by -\histoymin
    \histoy=\histoheight
    \divide\histoy by \histoydif
    \histoxu=\histox \histoyu=\histoy
    \ignorespaces}
\def\endhistogram{\kern\histowidth\hskip 0pt plus 0.5fil
    \advance\totalhistowidth\histowidth}
\def\histoxross(#1+#2-#3,#4+#5-#6){%
    \histocross(#1+#2-#3,#4+#5-#6)\histopoint(#1,#4)}
\def\histoaxes(+#1-#2,+#3-#4)#5#6{\histocross(0+#1-#2,0+#3-#4)
    \histoput(#1,0){\llap{\kern-10truept\linefont
    \char"2D}\lower 0.5ex\hbox{\ #5}}
    \histoput(0,#3){\lower 10truept\hbox to \z@{\linefont
    \char"36\hss}\raise 1ex\hbox to \z@{\hss #6\hss}}}
\def\histocross(#1+#2-#3,#4+#5-#6){%
    \hbox to \z@{%
    {\histoxx=\histox \dimen@i=\histox
     \def\tmp@{#2}\ifx\tmp@\empty\def\tmp@{#3}\fi
     \count@=#1\advance\count@ by -\histoxmin
     \advance\count@ by -#3\multiply\histox by \count@
     \kern\histox
     \count@=#4\advance\count@ by -\histoymin
     \multiply\histoy by \count@
     \advance\histoy by -0.5\histobarwidth
     \dimen@=\histoy \advance\dimen@ by \histobarwidth
     \count@@=#3 \advance\count@@ by \tmp@
     \multiply\histoxx by \count@@
     \ifhistopolyline\advance\histoxx by 0.25\histobarwidth\fi
     \vrule height \dimen@ depth -\histoy width \histoxx
     \multiply\dimen@i by \tmp@
     \ifhistopolyline\advance\dimen@i by 0.75\histobarwidth
     \else
     \advance\dimen@i by 0.5\histobarwidth
     \fi
     \kern -\dimen@i}%
    {\histoyy=\histoy
     \count@=#4\advance\count@ by -\histoymin
     \count@@=\count@
     \def\tmp@{#5}\ifx\tmp@\empty\def\tmp@{#6}\fi
     \advance\count@ by -#6
     \advance\count@@ by \tmp@
     \multiply\histoyy by \count@@
     \multiply\histoy by \count@
     \ifhistopolyline\advance\histoyy by 0.25\histobarwidth\fi
     \vrule height \histoyy depth -\histoy width \histobarwidth}%
    \hss}\ignorespaces}
\def\histohline(#1+#2-#3,#4){%
    \hbox to \z@{%
    {\histoxx=\histox \dimen@i=\histox
     \def\tmp@{#2}\ifx\tmp@\empty\def\tmp@{#3}\fi
     \count@=#1\advance\count@ by -\histoxmin
     \advance\count@ by -#3
     \multiply\histox by \count@
     \kern\histox
     \count@=#4\advance\count@ by -\histoymin
     \multiply\histoy by \count@
     \advance\histoy by -0.5\histobarwidth
     \dimen@=\histoy \advance\dimen@ by \histobarwidth
     \count@@=#3 \advance\count@@ by \tmp@
     \multiply\histoxx by \count@@
     \vrule height \dimen@ depth -\histoy width \histoxx
     \multiply\dimen@i by \tmp@
     \kern -\dimen@i \kern -0.5\histobarwidth}%
    \hss}\ignorespaces}
\def\histovline(#1,#2+#3-#4){%
    \hbox to \z@{%
    {\count@=#1\advance\count@ by -\histoxmin
     \multiply\histox by \count@
     \advance\histox by -0.5\histobarwidth
     \kern\histox
     \histoyy=\histoy
     \count@=#2\advance\count@ by -\histoymin
     \count@@=\count@
     \def\tmp@{#3}\ifx\tmp@\empty\def\tmp@{#4}\fi
     \advance\count@ by -#4
     \advance\count@@ by \tmp@
     \multiply\histoyy by \count@@
     \multiply\histoy by \count@
     \vrule height \histoyy depth -\histoy width \histobarwidth}%
    \hss}\ignorespaces}
\def\histopoint(#1,#2){%
    \hbox to \z@{%
    \count@=#1\advance\count@ by -\histoxmin
    \multiply\histox by \count@
    \kern\histox
    \count@=#2\advance\count@ by -\histoymin
    \multiply\histoy by \count@
    \raise\histoy\hbox to\z@{\hss\vbox to \z@{\vss
    \hbox{$\plotsymbol$}\vss}\hss}\hss}%
    \ignorespaces}
\def\histoput(#1,#2)#3{%
    \hbox to \z@{%
    {\count@=#1\advance\count@ by -\histoxmin
     \multiply\histox by \count@
     \kern\histox
     \count@=#2\advance\count@ by -\histoymin
     \multiply\histoy by \count@
     \raise\histoy\hbox to \z@{#3\hss}\hss}}%
     \ignorespaces}
\def\histomark#1{%
    \hbox to \z@{%
    {\kern\histowidth
     \histoy=\histoheight\advance\histoy by -3ex
     \raise\histoy\hbox to \z@{\hss#1\kern 0.9em}\hss}}%
     \ignorespaces}

\def\histoxxtick#1#2#3#4{%
    {\count@=#1\advance\count@ by -\histoxmin
     \multiply\histox by \count@
     \advance\histox by -0.5\histotickwidth
     \hbox to \z@{\kern\histox
     \vrule height #2 width \histotickwidth \hss}%
     \hbox to \z@{\kern\histox
     \ifdim#3=\z@\else
     \dimen@=\histoheight \advance \dimen@ by -#3
     \vrule height \histoheight depth -\dimen@ width \histotickwidth\fi
     \hbox{\vtop{\normalbaselines\halign{&##\cr\cr
     \hbox to \z@{\hss\ignorespaces#4\hss}\cr}}}%
     \hss}%
    }\ignorespaces}
\def\histoytick#1#2#3{%
    {\count@=#1\advance\count@ by -\histoymin
     \multiply\histoy by \count@
     \advance\histoy by -0.5\histotickwidth
     \dimen@=\histoy \advance\dimen@ by \histotickwidth
     \hbox to \z@{%
     \vrule height \dimen@ depth -\histoy width #2\hss}%
     \hbox to \z@{\kern\histowidth \kern-#2
     \vrule height \dimen@ depth -\histoy width #2\hss}%
     \advance\histoy by -0.5ex
     \raise\histoy\hbox to \z@{\hss #3\kern 0.4em}%
    }\ignorespaces}
\def\histoyltick#1#2#3{%
    {\count@=#1\advance\count@ by -\histoymin
     \multiply\histoy by \count@
     \advance\histoy by -0.5\histotickwidth
     \dimen@=\histoy \advance\dimen@ by \histotickwidth
     \hbox to \z@{%
     \vrule height \dimen@ depth -\histoy width #2\hss}%
     \advance\histoy by -0.5ex
     \raise\histoy\hbox to \z@{\hss #3\kern 0.4em}%
    }\ignorespaces}
\def\histoyrtick#1#2#3{%
    {\count@=#1\advance\count@ by -\histoymin
     \multiply\histoy by \count@
     \advance\histoy by -0.5\histotickwidth
     \dimen@=\histoy \advance\dimen@ by \histotickwidth
     \hbox to \z@{\kern\histowidth \kern-#2
     \vrule height \dimen@ depth -\histoy width #2\hss}%
     \advance\histoy by -0.5ex
     \raise\histoy\hbox to \z@{\kern\histowidth\kern 0.4em #3\hss}%
    }\ignorespaces}
\def\plotverythick{%
    \plotpointradius=38\p@ \divide\plotpointradius by 50
    \plotdotspacing=1.4\plotpointradius
    \setbox\plotpointbox\hbox to \z@{\hss\bigrm.\hss}%
    \ignorespaces}
\def\plotthick{%
    \plotpointradius=38\p@ \divide\plotpointradius by 72
    \plotdotspacing=1.4\plotpointradius
    \setbox\plotpointbox\hbox to \z@{\hss\tenrm.\hss}%
    \ignorespaces}
\def\plotnormal{%
    \plotpointradius=29\p@ \divide\plotpointradius by 72
    \plotdotspacing=1.4\plotpointradius
    \setbox\plotpointbox\hbox to \z@{\hss\sevenrm.\hss}%
    \ignorespaces}
\def\plotthin{%
    \plotpointradius=22\p@ \divide\plotpointradius by 72
    \plotdotspacing=1.4\plotpointradius
    \setbox\plotpointbox\hbox to \z@{\hss\fiverm.\hss}%
    \ignorespaces}
\def\plotline#1#2#3#4{%
    {\dimen@i=#1\dimen@ii=#2\dimen@@=#3\dimen@@i=#4\relax
    \advance\dimen@@ by -\dimen@i\advance\dimen@@i by -\dimen@ii
    \ifdim\dimen@@<\z@ \dimen@v=-\dimen@@ \else \dimen@v=\dimen@@ \fi
    \ifdim\dimen@@i<\z@ \dimen@=-\dimen@@i \else \dimen@=\dimen@@i \fi
    \ifdim\dimen@>\dimen@v \dimen@vi=\dimen@v \dimen@v=\dimen@ \else
    \dimen@vi=\dimen@ \fi
    \ifdim\dimen@vi>0.1\dimen@v
    \dimen@=10\dimen@v \divide\dimen@ by \dimen@vi \count@=\dimen@
    \divide \dimen@vi by \count@ \advance\dimen@v by 4.14\dimen@vi \fi
    \advance\dimen@ii by -\plotpointradius
    \ifdim\dimen@v>\plotdotspacing
      \advance\dimen@v by 0.75\plotdotspacing
      \divide\dimen@v by \plotdotspacing  
      \count@=\dimen@v
      \divide\dimen@@ by \count@ \divide \dimen@@i by \count@
      \loop
      \raise\dimen@ii\hbox to \z@{\kern\dimen@i\copy\plotpointbox\hss}%
      \advance\dimen@i by \dimen@@
      \advance\dimen@ii by \dimen@@i
      \advance\count@ by -1
      \ifnum\count@ > 0 \repeat
    \else
      \raise\dimen@ii\hbox to \z@{\kern\dimen@i\copy\plotpointbox\hss}%
    \fi}\ignorespaces}
\def\histolinee(#1,#2)(#3,#4){\histoline(#1,#2)(#3,#4)\histodraw(#3,#4)}
\def\histoline(#1,#2)(#3,#4){%
    \global\histoxx=\histox \global\histoyy=\histoy
    {\count@=#1\advance\count@ by -\histoxmin
     \multiply\histox by \count@
     \count@=#3\advance\count@ by -\histoxmin
     \global\multiply\histoxx by \count@
     \count@=#2\advance\count@ by -\histoymin
     \multiply\histoy by \count@
     \count@=#4\advance\count@ by -\histoymin
     \global\multiply\histoyy by \count@
     \plotline{\histox}{\histoy}{\histoxx}{\histoyy}}\ignorespaces}
\def\histomove(#1,#2)#3\histodraw#4({\histoline(#1,#2)(}
\def\histodraw(#1,#2){%
    {\count@=#1\advance\count@ by -\histoxmin
     \multiply\histox by \count@
     \count@=#2\advance\count@ by -\histoymin
     \multiply\histoy by \count@
     \plotline{\histoxx}{\histoyy}{\histox}{\histoy}
     \global\histoxx=\histox \global\histoyy=\histoy
    }\ignorespaces}
\def\histoascend(#1,#2)(#3){%
    \global\histoxx=\histox \global\histoyy=\histoy
    {\count@=#1\advance\count@ by -\histoxmin
     \multiply\histox by \count@
     \count@=#3\advance\count@ by -\histoxmin
     \global\multiply\histoxx by \count@
     \count@=#2\advance\count@ by -\histoymin
     \multiply\histoy by \count@
     \plotascend{\histox}{\histoy}{\histoxx}}\ignorespaces}
\def\plotascend#1#2#3{%
    {\dimen@i=#1\dimen@ii=#2\dimen@@=#3\relax
    \advance\dimen@@ by -\dimen@i
    \ifdim\dimen@@<\z@ \dimen@v=-\dimen@@ \else \dimen@v=\dimen@@ \fi
    \dimen@@i = \t@ntruept
    \ifdim\dimen@v>\dimen@@i
      \divide\dimen@v by \dimen@@i   
      \count@=\dimen@v \count@@=\count@ \advance\count@ by 1
      \advance\dimen@@ by -\count@\dimen@@i \divide\dimen@@ by \count@@
      \advance\dimen@@ by \dimen@@i
      \loop
      \raise\dimen@ii\hbox to \z@{\kern\dimen@i\linefont\char0\hss}%
      \advance\dimen@i by \dimen@@
      \advance\dimen@ii by \dimen@@
      \advance\count@ by -1
      \ifnum\count@ > 0 \repeat
    \fi}\ignorespaces}
\def\histodescend(#1,#2)(#3){%
    \global\histoxx=\histox \global\histoyy=\histoy
    {\count@=#1\advance\count@ by -\histoxmin
     \multiply\histox by \count@
     \count@=#3\advance\count@ by -\histoxmin
     \global\multiply\histoxx by \count@
     \count@=#2\advance\count@ by -\histoymin
     \multiply\histoy by \count@
     \plotdescend{\histox}{\histoy}{\histoxx}}\ignorespaces}
\def\plotdescend#1#2#3{%
    {\dimen@i=#1\dimen@ii=#2\dimen@@=#3\relax
    \advance\dimen@@ by -\dimen@i
    \ifdim\dimen@@<\z@ \dimen@v=-\dimen@@ \else \dimen@v=\dimen@@ \fi
    \dimen@@i = \t@ntruept
    \ifdim\dimen@v>\dimen@@i
      \divide\dimen@v by \dimen@@i   
      \count@=\dimen@v \count@@=\count@ \advance\count@ by 1
      \advance\dimen@@ by -\count@\dimen@@i \divide\dimen@@ by \count@@
      \advance\dimen@@ by \dimen@@i
      \advance\dimen@ii by -\dimen@@i
      \loop
      \raise\dimen@ii\hbox to \z@{\kern\dimen@i\linefont\char64\hss}%
      \advance\dimen@i by \dimen@@
      \advance\dimen@ii by -\dimen@@
      \advance\count@ by -1
      \ifnum\count@ > 0 \repeat
    \fi}\ignorespaces}

\def\plotsample#1{{\hbox{#1%
    \histowidth=6.9truemm \advance \histowidth by \plotdotspacing
    \plotline{0mm}{0.6ex}{\histowidth}{0.6ex}\kern 7truemm}}}
\catcode`\@=12 
\plotnormal
\def\startfigb#1{\startfigbody\kern #1\advance\totalhistowidth #1}
\expandafter\ifx\csname startfigbody\endcsname\relax%
\long\def\startfigbody{\vskip 0pt\hbox to \hsize\bgroup
     \vrule width \figrulewidth\hskip\figbodyleft
     \totalhistowidth=\figbodyleft
     \hskip 0pt plus 0.5fil\relax}%
\long\def\endfigbody{\hskip 0pt plus 0.5fil\vrule width \figrulewidth\egroup}
\def\figbodyleft{0pt}
\fi
\expandafter\ifx\csname figrulewidth\endcsname\relax%
\def\figrulewidth{0pt}\fi

\long\def\fighcapbox#1{\hfil
    \lower 1.4ex\hbox{\vbox{\advance\hsize -\totalhistowidth
    \advance\hsize -1.6em
    \hrule width \hsize  height 0pt
    #1
    }}}

\def\currentcolor{0 0 0;}
\def\startcolor#1{\setcolor{#1}%
 \xdef\currentcolor{#1;\currentcolor}\ignorespaces}
\def\setcolor#1{%
\expandafter\ifx\csname pdfoutput\endcsname\relax
 \else
 \ifnum\pdfoutput=1
 \else 
\fi\fi}
\def\getcolor#1;#2;#3\endget{\gdef\currentcolor{#2;#3}\gdef\oldcolor{#2}}
\def\endcolor{%
 \expandafter\getcolor\currentcolor\endget
 \expandafter\setcolor\expandafter{\oldcolor}%
 \ignorespaces}
\def\red{\startcolor{1 0 0}}
\def\Red#1{\red{#1}\endcolor{}}

\def\blue{\startcolor{0 0 1}}

\def\Blue#1{\blue{#1}\endcolor{}}

\def\de{\,{\rm d}}

\def\Ham{{\cal H}}

\def\Melbar{{\cal\kern 0.3em\overline{\kern -0.3em M\kern -0.1em}%
    \kern 0.1em}{}}
\def\Y#1S{{\Upsilon(#1{\rm S})}}
\def\subdecay{\hbox{\vrule height 7pt depth -2pt
    \kern -1pt\tensy\char"21}}
\def\M#1#2#3{\ifmmode{}^{#1}{\rm #2}_{#3}\else
    \hbox{${}^{#1}{\rm #2}_{#3}$}\fi}

\def\Ks{{K^0_{\scriptscriptstyle S}}}
\def\Kl{{K^0_{\scriptscriptstyle L}}}

\def\Jpsi{{J\mskip -3mu/\mskip -2mu\psi\mskip 2mu}}

\def\parall{{\mathchoice{\parallel}{\parallel}
    {\hbox{\kern 0.2ex\vrule height 1.1ex depth 0pt
    \vdistance{height 1.3ex depth 0.1ex}%
    \kern 0.21ex\vrule height 1.1ex depth 0pt\kern 0.2ex}}
    {\hbox{\kern 0.18ex\vrule height 0.8ex depth 0pt
    \vdistance{height 0.90ex depth 0.1ex}%
    \kern 0.18ex\vrule height 0.8ex depth 0pt\kern 0.18ex}}}}

\mathchardef\Gamma="7100
\def\Gam{{\rm\Gamma}}
\def\Del{{\rm\Delta}}
\mathchardef\Delta="7101
\mathchardef\Theta="7102
\mathchardef\Lambda="7103
\mathchardef\Xi="7104
\mathchardef\Pi="7105
\mathchardef\Sigma="7106
\mathchardef\Upsilon="7107
\mathchardef\Phi="7108
\mathchardef\Psi="7109
\mathchardef\Omega="710A
\def\qbar{\bar q}
\def\ubar{\bar u}

\def\Kbar{{\kern 0.2em\overline{\kern -0.2em K}}{}}
\def\Bbar{{\kern 0.18em\overline{\kern -0.18em B}}{}}
\def\Dbar{{\kern 0.2em\overline{\kern -0.2em D}}{}}
\def\Xbar{{\kern 0.2em\overline{\kern -0.2em X}}{}}
\def\anu{{\kern 0.06em\overline{\kern -0.06em \nu}}{}}
\def\pbar{{\kern 0.06em\overline{\kern -0.06em p}}{}}
\def\nbar{{\kern 0.06em\overline{\kern -0.06em n}}{}}
\def\Nbar{{\kern 0.2em\overline{\kern -0.2em N}}{}}
\def\Deltabar{{\kern 0.25em\overline{\kern -0.25em \Delta}}{}}
\def\Lbar{{\kern0.2em\overline{\kern-0.2em\Lambda\kern0.05em}%
    \kern-0.05em}{}}
\def\Sbar{{\kern 0.2em\overline{\kern -0.2em \Sigma}}{}}
\def\Xibar{{\kern 0.2em\overline{\kern -0.2em \Xi}}{}}
\def\Obar{{\kern 0.2em\overline{\kern -0.2em \Omega}}{}}
\let\Unitfont\rm
\def\Unit#1{\mskip 2mu{\Unitfont #1}\mskip 1mu}

\def\eV{{e\kern -0.08em V}}

\def\GeV{\Unit{G\eV}}

\def\Abar{{\kern 0.2em\overline{\kern -0.2em A\kern -0.03em}\kern 0.03em}{}}
\def\<#1#2{#1{\phantom{#2}}}

\begin{document}

\title{B MESON DECAYS AND CP (AND T) VIOLATION
}

\author{Roland Waldi}

\address{Institut f\"ur Physik\\
Universit\"at Rostock \\
August-Bebel-Str. 55, 18055 Rostock, Germany\\
E-mail: roland.waldi@uni-rostock.de }

\maketitle

\abstracts{The prolific field of $B$ meson decays and CP violation
is illustrated in a few examples of recent results:
the measurement of the CKM unitarity angle $\beta = \phi_1$,
the measurement of a significant violation of time reversal symmetry,
an unexplained isospin asymmetry in penguin decays,
a hint on scalar charged bosons from
the semileptonic $B$ decay to the heavy lepton $\tau$,
and $B$ decays to baryons.
}

\section{Introduction}

The B Factories PEP~II/BABAR and KEK-B/Belle have been
constructed at the end of the last century
to search for CP violation in the decays of $B$ mesons.
The discovery of CP asymmetries has confirmed the
Standard Model picture and led to a Nobel prize for
Kobayashi and Maskawa, the inventors of a 3-family quark mixing matrix
\cite{ckm}.
Both collaborations are still producing new results on $B$ meson
CP asymmetries and many other features of $B$ meson decays.
While the B Factories produce $B^+ B^-$ and $B^0 \Bbar^0$ pairs
exclusively
from decays of the $\Y4S$ formed in $e^+ e^-$ annihilation,
the LHCb detector at the $p p$ collider LHC has recently appeared
on this stage, producing much higher numbers of $B$ mesons and also
$b$ baryons buried in jets. They can be isolated by a highly performant
vertex detector.

\begin{figure}[tb]
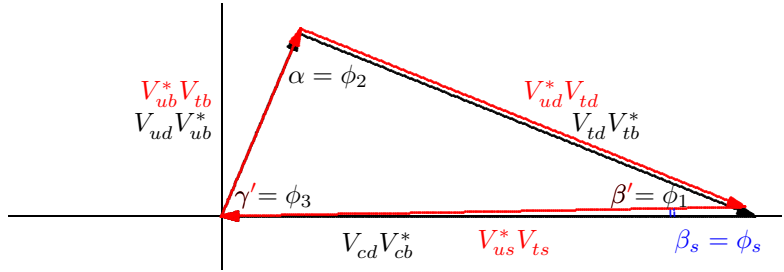

\histosize{0.84\hsize}{0.28\hsize}
\startfigb{0mm}\sf
\historulewidth=0pt
\starthistogram {-450}{1050} {-100}{400}
\histocross(0+1050-400,0+400-100)
\plotthick
\histoline (0,0)(1000,0)
 \histodraw(968,4) 
 \histodraw(974,19)
 \histodraw(1000,0)
 \histodraw(970,6)
 \histodraw(974,16)
 \histodraw(1000,0)
 \histodraw(970,9)
 \histodraw(972,14)
 \histodraw(1000,0) 
\histodraw      (145,343)
 \histodraw(141,311) 
 \histodraw(125,318)
 \histodraw(145,343)
 \histodraw(138,313)
 \histodraw(128,317)
 \histodraw(145,343)
 \histodraw(135,314)
 \histodraw(131,316)
 \histodraw(145,343) 
\histodraw      (0,0)
 \histodraw(31,8) 
 \histodraw(31,-8)
 \histodraw(0,0)
 \histodraw(30,5)
 \histodraw(30,-5)
 \histodraw(0,0)
 \histodraw(31,3)
 \histodraw(31,-3)
 \histodraw(0,0) 
\Red{%
\plotthick
\histoline (0,0)(982,17)
 \histodraw(950,21) 
 \histodraw(956,36)
 \histodraw(982,17)
 \histodraw(952,23)
 \histodraw(956,33)
 \histodraw(982,17)
 \histodraw(952,26)
 \histodraw(954,31)
 \histodraw(982,17) 
\histodraw      (149,352)
 \histodraw(145,320) 
 \histodraw(129,327)
 \histodraw(149,352)
 \histodraw(142,322)
 \histodraw(132,326)
 \histodraw(149,352)
 \histodraw(139,323)
 \histodraw(135,325)
 \histodraw(149,352) 
\histodraw      (0,0)
 \histodraw(31,9) 
 \histodraw(31,-8)
 \histodraw(0,0)
 \histodraw(30,6)
 \histodraw(31,-5)
 \histodraw(0,0)
 \histodraw(31,3)
 \histodraw(31,-2)
 \histodraw(0,0) 
\histoput (730,25){$\beta'$}
\histoput (25,25){$\gamma'$}
\histoput (480,0) {\lower 12pt\hbox{$V_{u s}^* V_{t s}\<^*
           $}}
\histoput (550,220) {\ $V_{u d}^* V_{t d}\<^*
           $}
\histoput (0,160) {\raise 12pt\hbox to 0pt{\hss$V_{u b}^* V_{t b}\<^*
           $\ }}
}%
\Blue{\plotthin
\histoline(850,0)(849,15)
\histoline(843,0)(842,14)
\histoput (850,0) {\lower 12pt\hbox{$\beta_s = \phi_s$}}}%
\histoput (730,25){$\beta = \phi_1$}
\histoput (125,250){$\alpha = \phi_2$}
\histoput (25,25){$\gamma = \phi_3$}
\histoput (225,-5) {\lower 12pt\hbox to 0pt{$V_{c d}\<^* V_{c b}^*
           $\hss}}
\histoput (640,160) {\ $V_{t d}\<^* V_{t b}^*
           $}
\histoput (0,160) {\hbox to 0pt{\hss$V_{u d}\<^* V_{u b}^*
           $\ }}
\endhistogram
\endfigbody
\caption[*]{Two of the six unitarity triangles of the
CKM matrix.  CP violation in $B$ decays
is parameterised via the angles (phase
differences) of these triangles.}
\label{fig:unitri}
\end{figure}

The explanation of CP violation in the Standard Model
is a physical phase in the CKM (Cabibbo-Kobayashi-Maskawa) matrix,
since all CP violating effects are interference phemomena.
However, the CKM matrix has 5 more phases, that can be changed
as arbitrary phases of quark fields.  Hence, the physical CKM phase
cannot be uniquely associated with specific matrix elements.
Two of the infinite number of completely equivalent
parametrisations of the CKM matrix are
\begin{equation}
V =
 \pmatrix {|V_{u d}| &|V_{u s}| &|V_{u b}| e^{-i\tilde\gamma}\cr
   -|V_{c d}| e^{i\tilde\phi_4} &|V_{c s}| e^{-i\tilde\phi_6} & |V_{c b}|\cr
    |V_{t d}| e^{-i\tilde\beta}&-|V_{t s}| e^{i\tilde\beta_s}& |V_{t b}|\cr }
\label{eq:ckm}
\end{equation}
adopted by the particle data group \cite{pdg}
where the large phases are attached to the smallest matrix elements,
and
$$
V =
\pmatrix {-|V_{u d}|e^{-i\alpha}
 &|V_{u s}|e^{i\tilde\gamma} &|V_{u b}| \cr
       -|V_{c d}| e^{i(\tilde\phi_4+\tilde\beta)}
        &|V_{c s}| e^{-i\tilde\phi_6} & |V_{c b}|\cr
        |V_{t d}| &-|V_{t s}| e^{i\tilde\beta_s}& |V_{t b}|\cr }
$$
where these elements are real and large phases are attached to
$V_{u d}$,
$V_{u s}$ and
$V_{c d}$.
Due to this arbitrariness, CP violation is parametrised via the angles
in the two non-flat unitarity triangles (Fig.\ \ref{fig:unitri})
which describe the row-1-3
and column-1-3 unitarity conditions in the complex plane.  The shapes
of unitarity triangles are independent of the phase
convention for the CKM matrix.

In the following sections, the lively and active field of $B$ decays and
CP violation will be illustrated in a few examples of recent
experimental results.
I will, however, not elaborate
on details of published analyses that can be found in the
original papers.

\section{Measurement of $\beta = \phi_1$}

The angle $\beta$ or $\phi_1$ is the angle measured via the first CP asymmetry
observation in
$B$ meson decays.
It is 
$\beta \approx \tilde\beta$ in Eq.~\ref{eq:ckm}, and has been
measured \cite{bab0118} via the decay to $\Jpsi \Ks$ and related decay 
channels.
The time-dependent CP asymmetry is
$$
\left.{\dot N(\Bbar \to X) - \dot N(B \to X) \over
        \dot N(\Bbar \to X) + \dot N(B \to X)}\right\vert_t
   = \sin 2\beta \sin \Del m\, t
$$
where $t$ at the B Factories is the difference of the decay times
of the two neutral $B$ mesons from the $\Y4S$ decay, and $B$ and $\Bbar$ designate
the flavour of the signal $B$ meson at the time of decay of the other
(tag) $B$ meson.

\begin{figure}[ht]
\centerline{
   \epsfxsize=0.25\columnwidth
   \epsfclipon
   \epsfbox{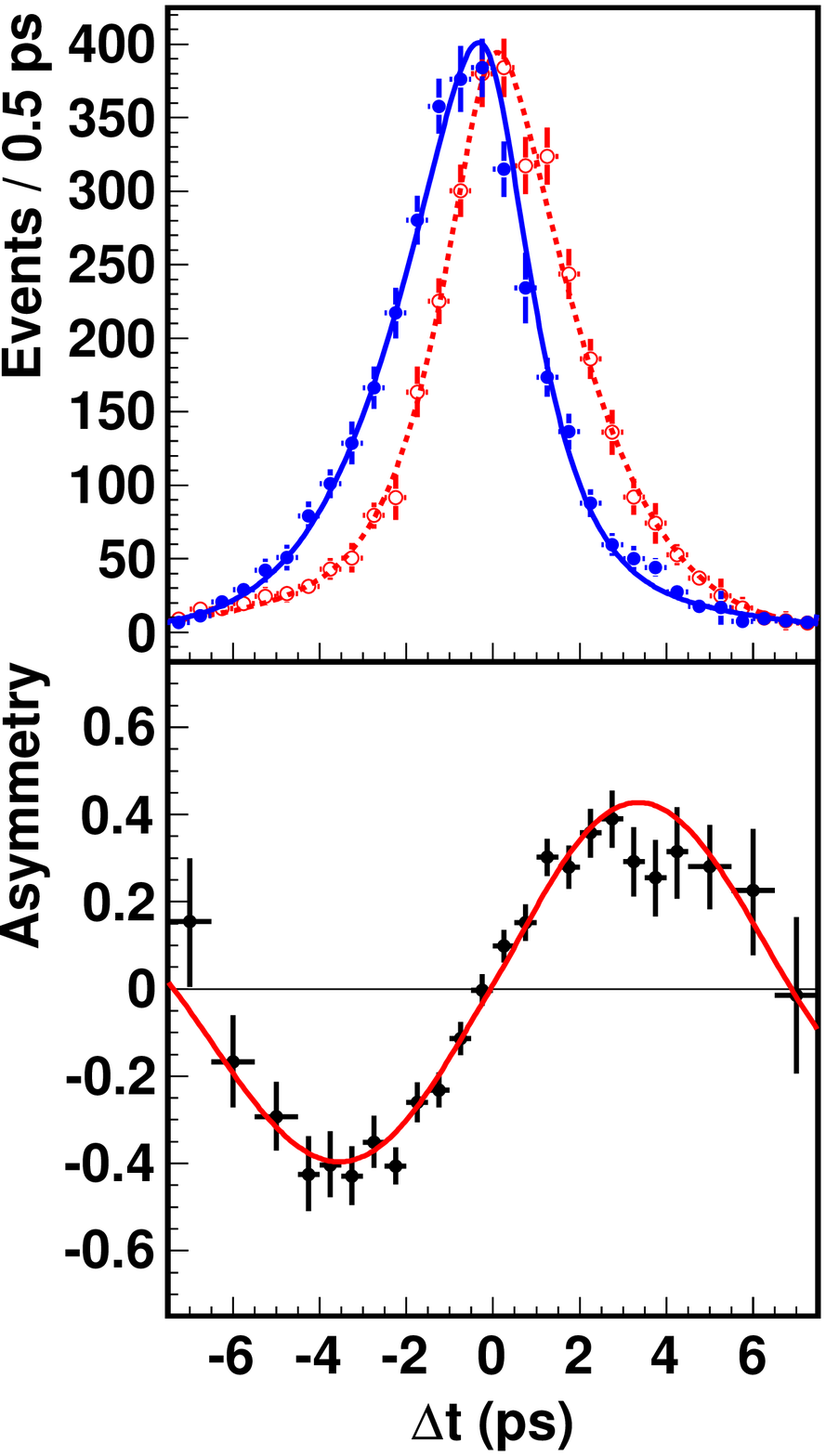}
\qquad
   \epsfxsize=0.25\columnwidth
   \epsfclipon
   \epsfbox{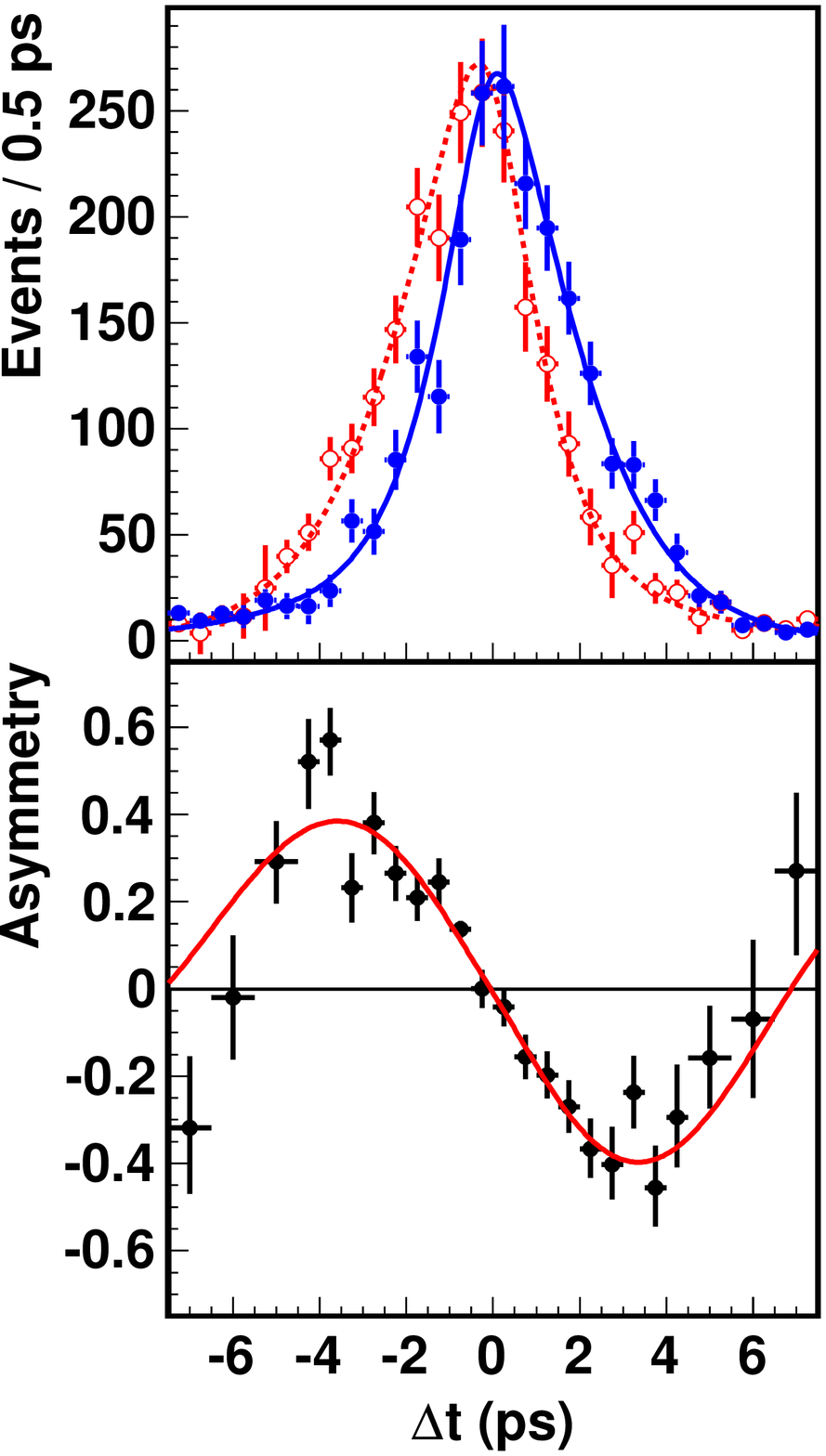}}
\caption[*]{The time dependent decay rates and CP asymmetries
from the measurement of $\phi_1 = \beta$ by Belle \protect\cite{bellebeta}
for CP $-1$ and CP $+1$ eigenstates.
}\label{Fig:bellebeta}
\end{figure}

The world average of $\beta$ is still dominated by the B factories.
The most precise single measurement has been recently
published by Belle \cite{bellebeta}.
Details of the analysis can be found in this publication.
The very clean decay rate and asymmetry plots are shown in Fig.\ \ref{Fig:bellebeta}.

The final states $D^{(*)+} D^{(*)-}$ show the same time behaviour
in their CP asymmetry when they are produced from the $b \to c$ tree
diagram.  However, the small amplitude from a $b \to d$ penguin
diagram can lead to direct CP violation and a modified asymmetry
amplitude.  The measurements by BABAR \cite{ddbabar}
and Belle \cite{ddbelle}, which marginally
disagree, do not yet help to fix this contribution.

\begin{figure}[tb]
\centerline{
   \epsfxsize=0.48\columnwidth
   \epsfbox{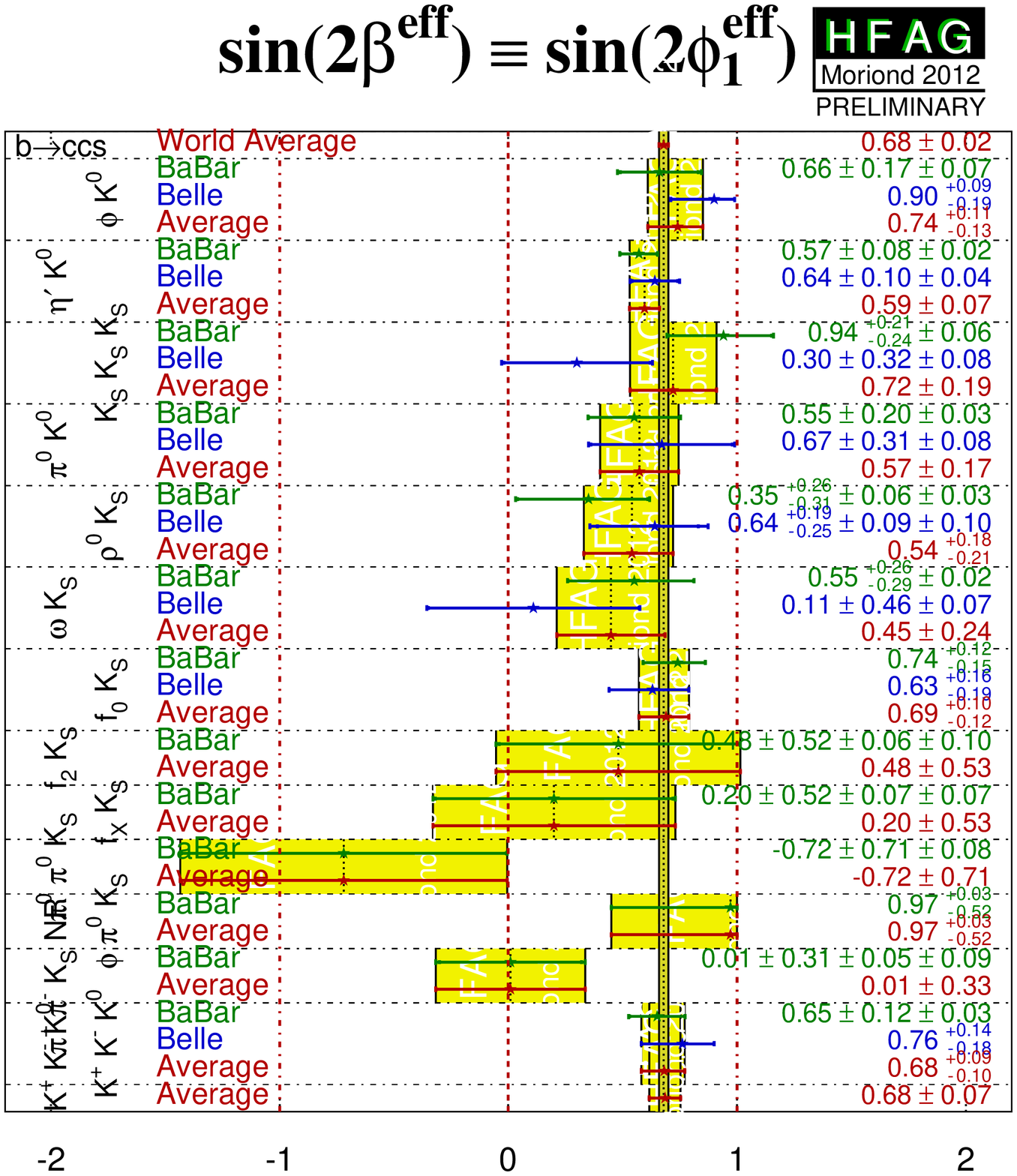}}
\caption[*]{The coefficients of the $\sin\Del m\,t$ CP asymmetry
for various penguin decays of the neutral $B$ meson, compared
to $\sin2\beta$ from the $\Jpsi\Ks$ final state.
}\label{Fig:betapeng}
\end{figure}

Decays proceeding via the $b \to s$ penguin diagram
show a CP asymmetry
$$
\left.{\dot N(\Bbar \to X) - \dot N(B \to X) \over
        \dot N(\Bbar \to X) + \dot N(B \to X)}\right\vert_t
   = \sin 2\beta' \sin \Del m\, t
$$
where the value
$\beta'-\beta \approx \beta_s$ is expected to be small and
can be directly measured in CP asymmetries in $B_s$ decays
($\beta_s \approx \tilde\beta_s$ in Eq.~\ref{eq:ckm}).
In the past, a large discrepancy between $\sin 2\beta$ from $\Jpsi\Ks$
and $\sin 2\beta'$ from various penguin decays has caused some
irritation.  However, present values at improved precision have been
compiled by the Heavy Flavour Averaging Group
\cite{hfag} and are shown in Fig.\ \ref{Fig:betapeng}. They are statistically compatible
with the expectation $\beta' \approx \beta$.

Deviations should be expected, however, from other decay diagrams
contributing in addition to the leading penguin.
These could also lead to direct CP violation.
The measurement of a $\cos \Del m\, t$ amplitude in neutral $B$ meson
CP asymmetries
is by now not conclusive
to confirm these contributions.
But it is verified by a direct CP asymmetry recently seen 
by LHCb \cite{lhcbcpvhhh} in charged $B$
meson decays into three charged hadrons.  They find
\begin{eqnarray}
A_{CP}(B^\pm \to K^\pm \pi^+\pi^-) &= +0.034\pm0.009\pm0.008\nonumber\\
A_{CP}(B^\pm \to K^\pm K^+K^-) &= -0.046\pm0.009\pm0.009\nonumber
\end{eqnarray}
at $2.8\sigma$ and $3.7\sigma$ significance.

\section{T Violation}

One of the recent highlights
is the violation of time reversal symmetry T observed by BABAR \cite{babartv}.
This measurement is closely connected to time-dependent
CP violation,
both by the CPT theorem and by the measurement itself.  It makes use
of distributions like the ones shown in Fig.\ \ref{Fig:bellebeta}.
It is based on the entanglement of the two neutral $B$
mesons from $\Y4S \to B^0\Bbar^0$ decay which allows to
compare the transitions between flavour eigenstates
and other superpositions of $B^0$
and $\Bbar^0$ with their time-reversed counterpart.

When the oscillation $B^0 \to \Bbar^0$ is time reversed
to $\Bbar^0 \to B^0$, this transformation is also
a CP transformation.  This is not the case for the
time reversal of $B^0 \to B_+$ into $B_+ \to B^0$,
and likewise
$B^0 \to B_-$,
$\Bbar^0 \to B_+$, and
$\Bbar^0 \to B_-$.
Here, $B_+$ and $B_-$ are
orthogonal states defined via their decay products.
The $B_+$ state decays
to the CP $+1$ final state $\Jpsi\Kl$, and the $B_-$ to
a CP $-1$ final state $\Jpsi\Ks$ plus some other channels with the same
quark content and CP eigenvalue.

CP eigenstates from $B^0$ and $\Bbar^0$ wave functions
cannot be constructed unambiguously due to the arbitrary quark phases,
but
the CP eigenvalue of the final state $\Jpsi(\pi^+\pi^-)_{\Ks}$ where
$\Ks$ decays to the $\pi^+\pi^-$ CP $+1$ eigenstate (this state
is denoted $\Jpsi\Ks_{\rm CP+}$ below) is uniquely defined
to be $-1$.
Defining the decay amplitudes
\begin{eqnarray}
A_- &= \langle \Jpsi\Ks_{\rm CP+} | \Ham | B^0\rangle\nonumber\\
\Abar_- &= \langle \Jpsi\Ks_{\rm CP+} | \Ham | \Bbar^0\rangle\nonumber
\end{eqnarray}
it can be
shown \cite{alvszy} that the choice
$$
|B_+\rangle = {\Abar_- |B^0\rangle - A_- |\Bbar^0\rangle
              \over \sqrt{|\Abar_-|^2 + |A_-|^2}}
$$
and
$$
|B_-\rangle = {\Abar_-^* |\Bbar^0\rangle + A_-^* |B^0\rangle
              \over \sqrt{|\Abar_-|^2 + |A_-|^2}}
$$
defines two orthogonal states $B_{\pm}$.
It is obvious that
$$
\langle \Jpsi\Ks_{\rm CP+} | \Ham | B_+\rangle \propto
\Abar_- A_- - A_- \Abar_- = 0,
$$
i.e., the state $B_+$ cannot decay into the CP $-1$ final state
$\Jpsi\Ks_{\rm CP+}$.
Taking the approximation that $\Kl \approx \Kl_{\rm CP-}$
is a CP eigenstate, which is violated by only a few per mille,
it can also be argued that
$\Jpsi\Kl$ cannot be reached from
$B_-$ but from the orthogonal state $B_+$.
For this to hold,
we have
to assume (and verify) that the decays to
$\Jpsi\Ks$ and $\Jpsi\Kl$
proceed through the same diagrams.
These final states are then tagging the states $B_+$ and $B_-$
at the time of their decay.

On the other hand, flavour tagging final states like
$D^{*+} l^- \anu$ determine the state $\Bbar^0$ at decay time, while
$D^{*-} l^+ \nu$ determines a $B^0$.

The entanglement of the two $B$ mesons from $\Y4S$ decay guarantees that
at the time a $B_+$ or $B^0$ is detected on one side,
the other must be a $B_-$ or $\Bbar^0$, respectively.  When this
latter meson decays, it can---depending on the decay channel---sometimes
be identified as one of the states mentioned. Then a transition in time
for this meson can be observed.

These transitions are related by discrete symmetries, for example
\def\1{\hbox{\ before\ }}
\begin{eqnarray}
B^0 \to B_+\ (\Bbar^0 \1 B_+) & {\buildrel \rm CP \over \longrightarrow} & \Bbar^0 \to B_+\ (B^0 \1 B_+)\nonumber\\
B^0 \to B_+\ (\Bbar^0 \1 B_+) & {\buildrel \rm T \over \longrightarrow} & B_+ \to B^0\ (B_- \1 B^0) \\
B^0 \to B_+\ (\Bbar^0 \1 B_+) & {\buildrel \rm CPT \over \longrightarrow} & B_+ \to \Bbar^0\ (B_- \1 \Bbar^0) \nonumber
\end{eqnarray}
Here $(X \1 Y)$ means that the $B$ meson that decays first is tagged
as $X$ and the $B$ meson that decays second is tagged as $Y$ and is the
one whose state transition with time is observed.

The details of the analysis and a table with all T, CP and CPT asymmetries
are found in \cite{babartv}.
The results are a significant CP and T violation, and no CPT
asymmetry within errors, in agreement with the Standard Model.

\section{$B$ Decays: a Penguin and New Physics}

Many searches for new physics have been conducted through $B$ meson
decays.
One example with a still unexplained asymmetry is the penguin decay
$B \to K^{(*)} l^+ l^-$.  These decays have been observed by BABAR
\cite{babarpeng}, Belle \cite{bellepeng}, CDF \cite{cdfpeng},
and recently LHCb \cite{lhcbpengiso}.
The LHCb result is the first to establish an
isospin asymmetry between $B^0 \to K^0 \mu^+ \mu^-$ and
$B^+ \to K^+ \mu^+ \mu^-$.
The isospin asymmetry is defined as
$$
A_I(q^2) = {\de\Gam(B^0 \to K^0 l l)/\de q^2 -
          \de\Gam(B^+ \to K^+ l l)/\de q^2 \over
          \de\Gam(B^0 \to K^0 l l)/\de q^2 +
          \de\Gam(B^+ \to K^+ l l)/\de q^2}
$$
and is calculated in bins of the $\mu^+ \mu^-$ invariant mass squared
$q^2$.
Details of the analysis can be found in \cite{lhcbpengiso}.

\begin{figure}[tb]
\centerline{
   \epsfxsize=0.50\columnwidth
   \epsfbox{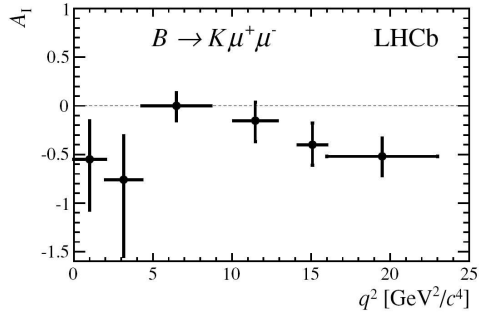}}
\caption[*]{Isospin asymmetry in $B \to K \mu^+ \mu^-$ at LHCb
as function of $q^2$, the $\mu^+ \mu^-$ invariant mass squared.
}\label{Fig:lhcbpengiso}
\end{figure}

Although the Standard Model has isospin
violating amplitudes like virtual photon emission by the light quark
of the $B$ meson or $b\ubar$ annihilation with triple boson
couplings, these amplitudes provide only very small differences
in the two decay rates.  The $4.4\sigma$ asymmetry shown
in Fig.\ \ref{Fig:lhcbpengiso} cannot be described in the Standard Model
and may be a first hint on New Physics, although no
explanation has been put forward by now.

The data seem to prefer charged $B$ meson decays over
neutral $B$ mesons.
There is a similar unexplained isospin asymmetry
in baryonic penguin decays, evaluated from present world
averages \cite{pdg}:
$$
{\Gamma(B^+ \to p\pbar K^+) \over
\Gamma(B^0 \to p\pbar K^0)} = 1.91\pm0.27
,\qquad
{\Gamma(B^+ \to p\pbar K^{*+}) \over
\Gamma(B^0 \to p\pbar K^{*0})} = 2.7\pm0.3
$$
which also favours charged $B$ meson decays.

\section{$B$ Decays: a Tree and New Physics}

While loop diagrams are generally notorious
for possible New Physics contributions,
the semileptonic decays to the heavy $\tau$ lepton is a tree diagram
that may show physics beyond the Standard Model.
In particular, a charged Higgs boson may become visible at
the high-mass $\tau$ lepton while its contribution
is negligible in the production of low-mass leptons.

The exclusive decays $B \to D^{(*)} \tau \nu$ have been
investigated in the past by
BABAR and Belle, but the results were compatible with the
Standard Model within precision.

The $B \to D^{(*)} \tau \nu$ decay has
more than two neutrinos in the final state and so cannot be fully
reconstructed using only the observable particles.  It therefore relies on
exclusive reconstruction of the accompanying $B$ meson to
provide the necessary level of background suppression.
A recent study by BABAR \cite{babardtaunu}
has not only improved the statistical precision, but has also
reduced systematic uncertainties using only leptonic $\tau$
decays, i.e., using the same detected states $D^{(*)} \mu$ and
$D^{(*)} e$ to reconstruct semileptonic decays to low-mass leptons and
to $\tau$.

The results show a marked discrepancy to
the Standard Model expectation at $3.4\sigma$.  This may be interpreted
as evidence for a charged Higgs boson, but predictions for a
type II two Higgs doublet model simulated by the experimentalists
cannot explain the results for $D$ and $D^*$ mesons simultaneously.
In a recent paper \cite{greubhiggs}, however, a possible explanation in
a type III two Higgs doublet model has been presented.

\section{$B$ Decays to Baryons}

Baryons in the final state occur in pairs due to baryon number conservation,
hence the minimum mass of a meson that can decay into baryons is
above $1.9\GeV$.  There are only very few examples at masses below the
$B$ meson:
$D_s^+ \to p\nbar$ and $\Jpsi \to {}$baryons.  But the weakly decaying
$B$ mesons with a mass of $5.28\GeV$ are far above the threshold and
allow a multitude of different channels, including, in principle,
even final states with four baryons.

\begin{figure}[t]
\centerline{\ a)\hfill\ b)\ \ \hfill}
\centerline{
   \epsfxsize=0.43\columnwidth
   \epsfbox{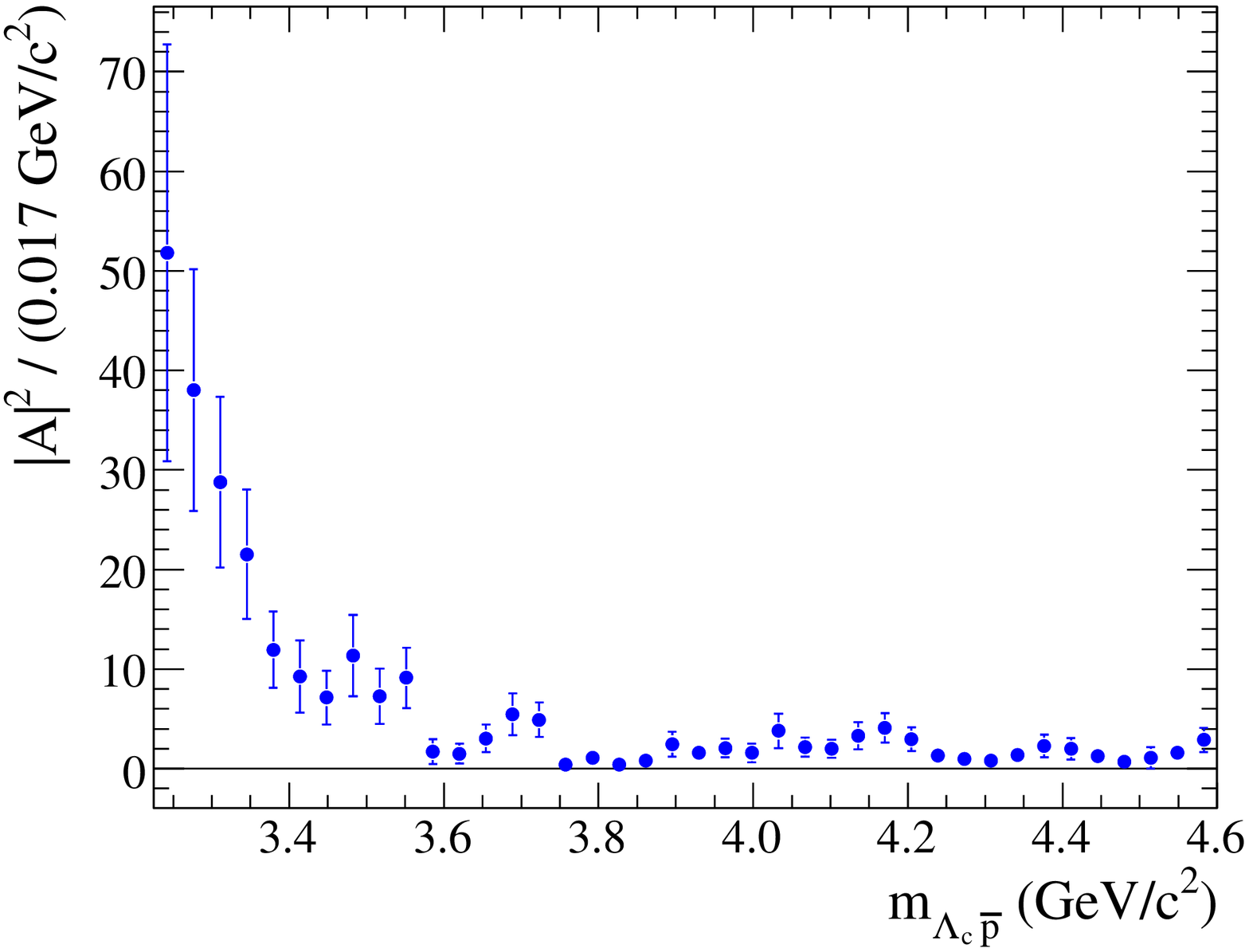}\hfill
   \epsfxsize=0.54\columnwidth
   \epsfbox{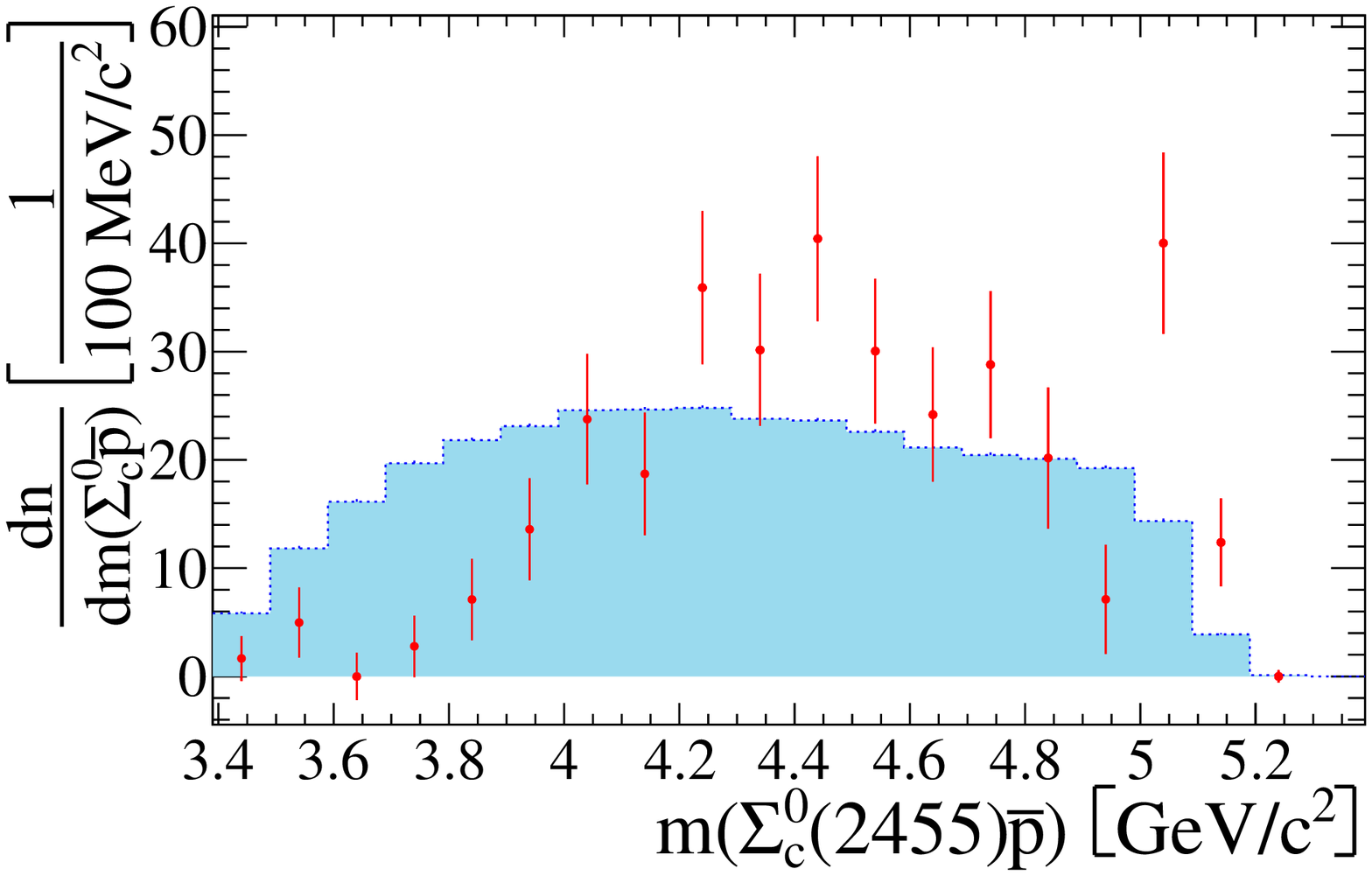}}
\caption[*]{Invariant baryon antibaryon mass distribution in the decay
(a) $B^- \to \Lambda_c^+ \pbar \pi^-$ divided by phase space
\protect\cite{Aubert2008if}, and
(b) $\Bbar^0\to\Sigma_{c}^{0}\pbar\pi^+$ (red points) compared
with a three-body phase space distribution (blue histogram).
}\label{Fig:barypairmass}
\end{figure}

Inclusive studies at ARGUS \cite{argus1992he}
have shown that $(6.8\pm0.6)\%$ of all $B$ decays have baryons in their
final state.  Exclusive channels have been investigated by CLEO
and the B Factories BABAR and Belle, adding up to
$(0.53\pm0.06)\%$ ($B^0$) and
$(0.85\pm0.15)\%$ ($B^+$), respectively \cite{pdg}.  This is still a small fraction
of the total, but enough to show some salient features:
two-body branching fractions are much smaller than similar three-body
rates, and there is still a moderate increase in branching fraction
to higher multiplicities.  Also, most multibody channels show a
more or less pronounced enhancement at the threshold of the
baryon anti-baryon invariant mass.

\begin{figure}[ht]
\scriptsize
\centerline{\ a)\hfill\ b)\hfill\ c)\hfill\ d)\hfill}
 \centerline{
   \epsfxsize=0.23\columnwidth
   \epsfbox{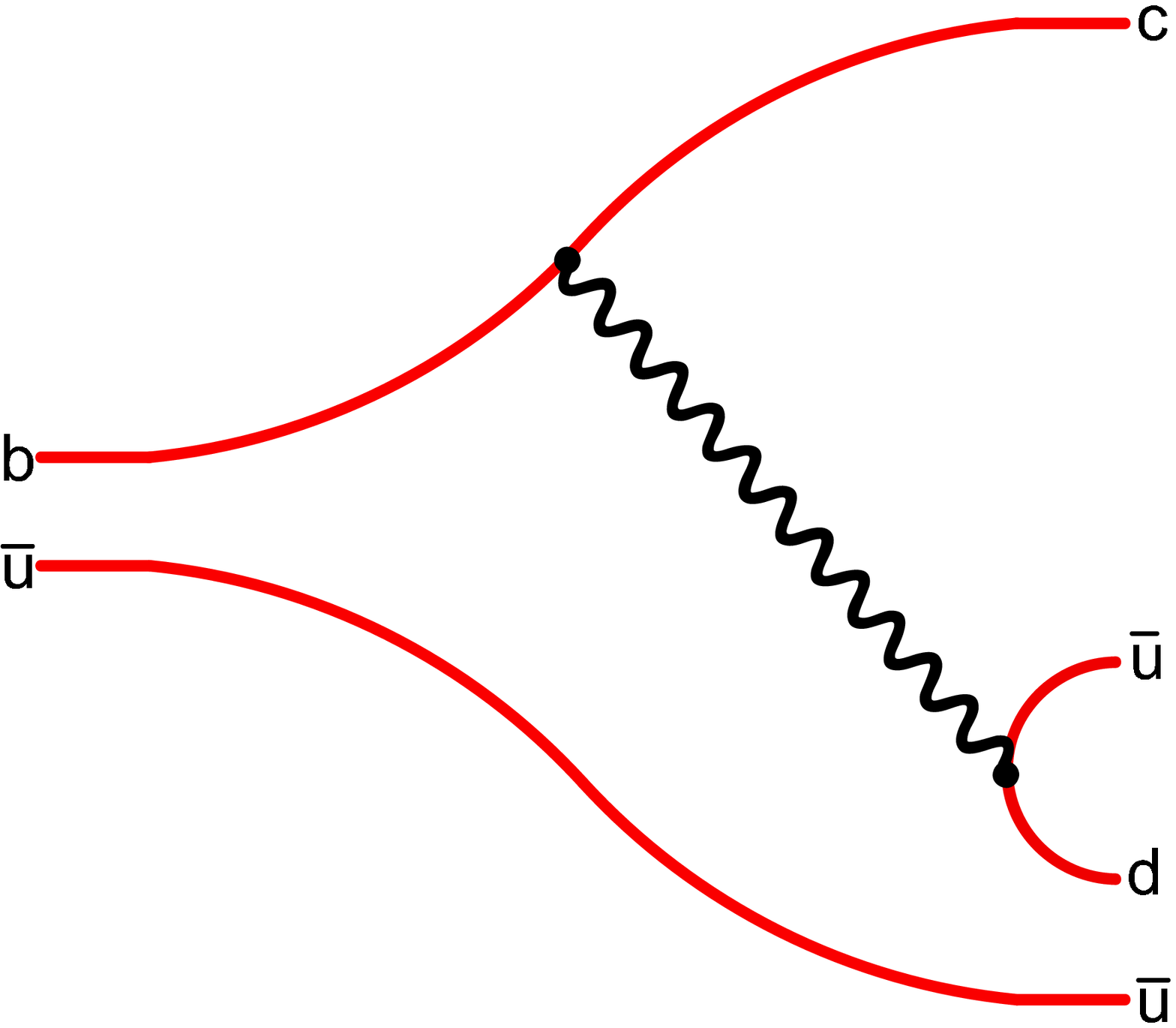}
\hfill
   \epsfxsize=0.23\columnwidth
   \epsfbox{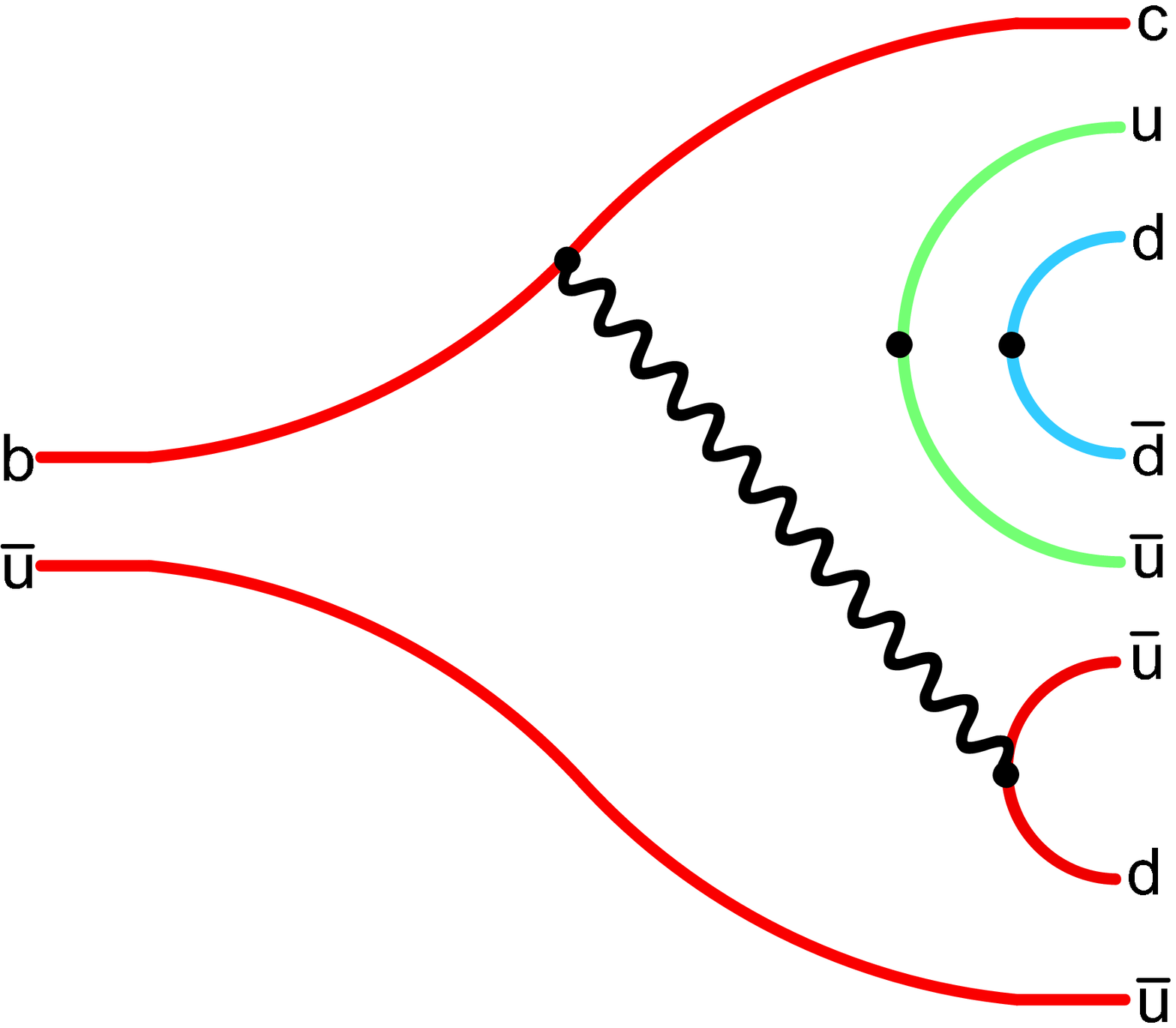}
\hfill
   \epsfxsize=0.23\columnwidth
   \epsfbox{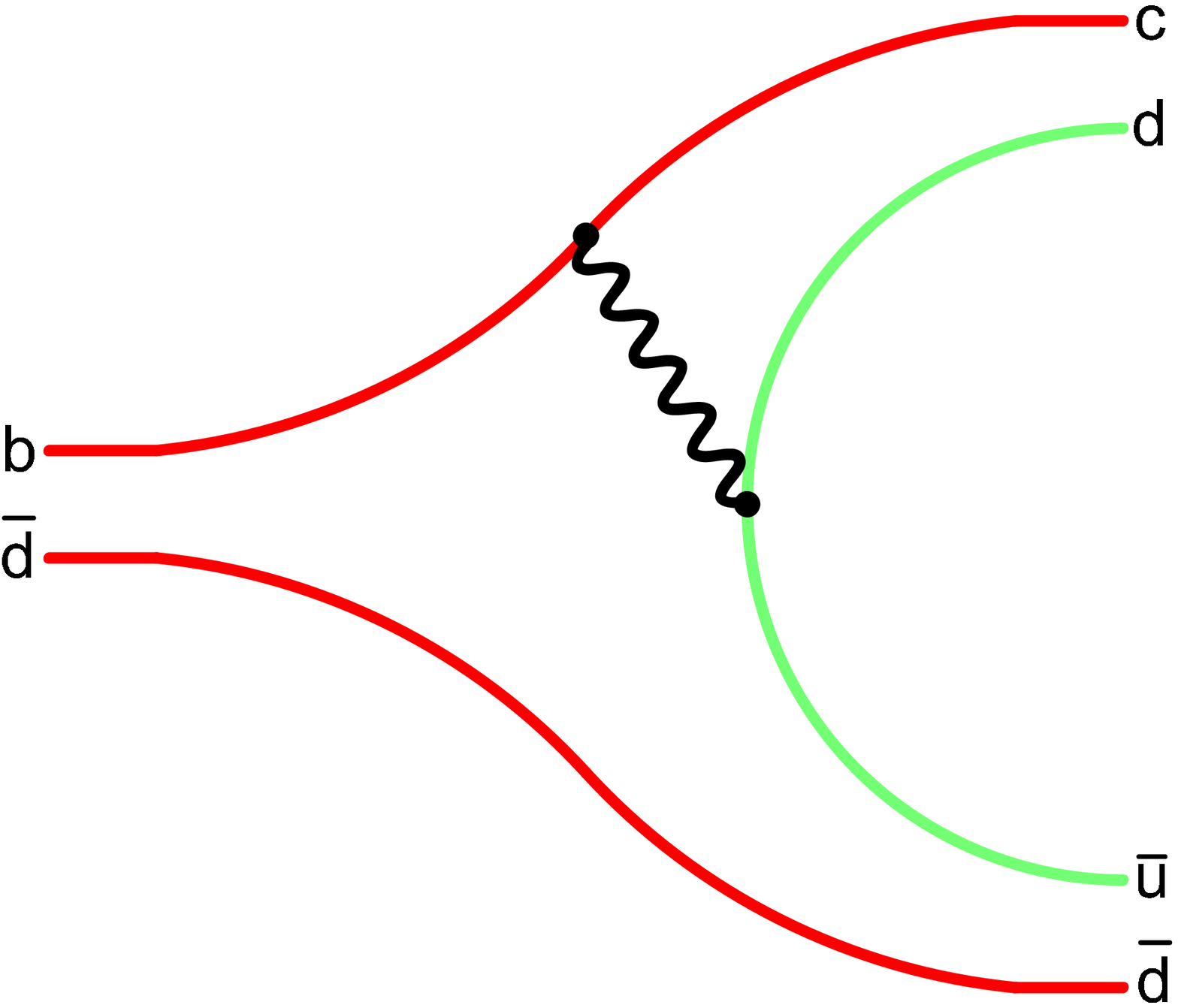}
\hfill
   \epsfxsize=0.23\columnwidth
   \epsfbox{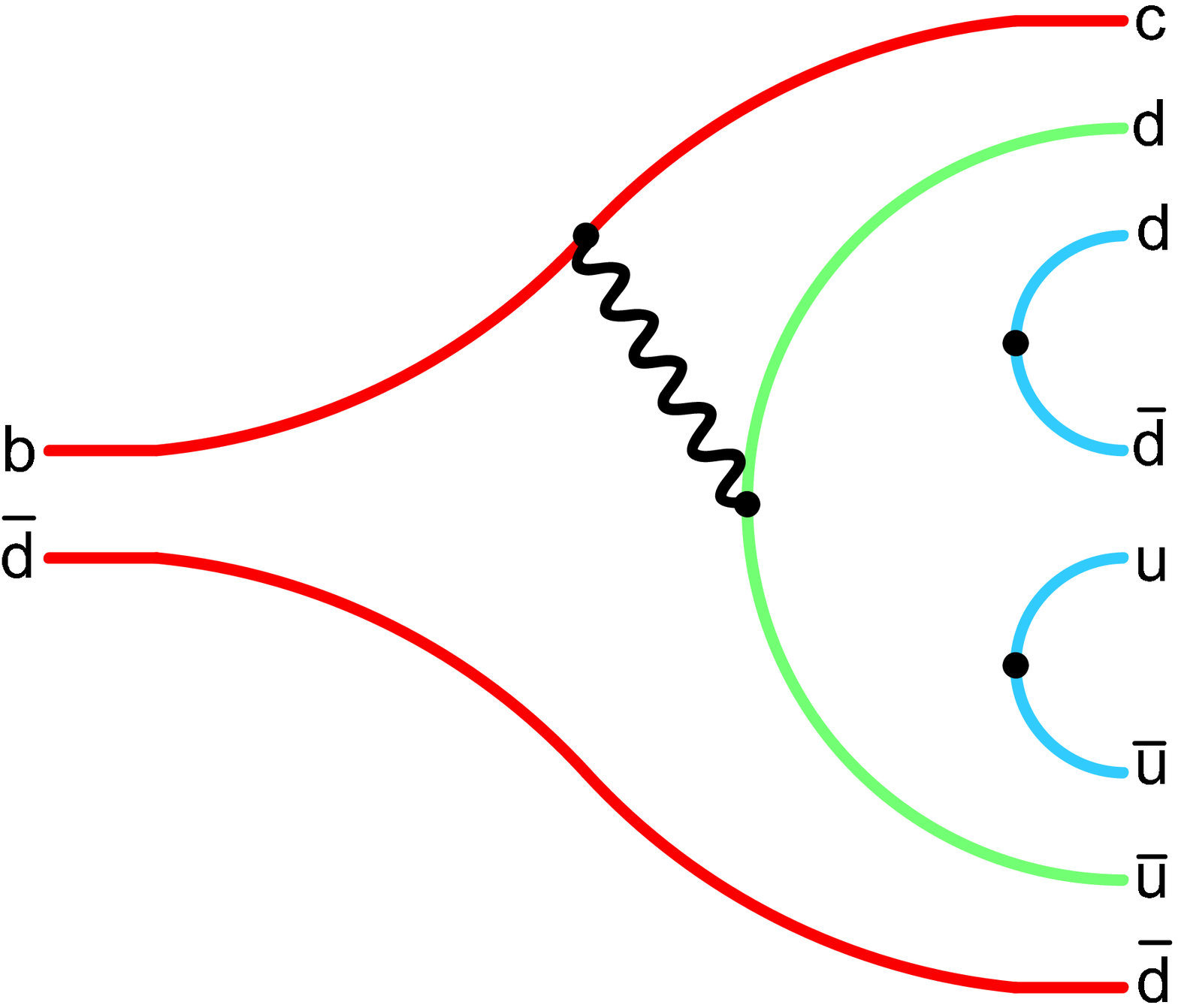}\hfill}
\caption[*]{Spectator diagrams (corresponding to operator $O_2$) illustrating the
    basic picture for a baryon antibaryon threshold enhancement:
    (b) shows one diagram for $B^- \to \Lambda_c^+ \pbar \pi^-$
       with preferentially low baryon antibaryon mass related to meson pair diagram (a), and
    (d) for $\Bbar^0\to\Sigma_{c}^{0}\pbar\pi^+$
       with preferentially high baryon antibaryon mass related to diquark pair diagram (c).
}\label{Fig:Bbaryfeyn}
\end{figure}

I want to elaborate a little bit on the latter feature, illustrated in
Fig.~\ref{Fig:barypairmass}a.
Examples of Feynman diagrams (where gluons are omitted) are shown
in Fig.~\ref{Fig:Bbaryfeyn}.  The colour suppressed diagram 
in Fig.~\ref{Fig:Bbaryfeyn}b 
as well as other tree and penguin diagrams
have one feature in common:  if we omit
the quarks created from the vacuum, they represent decays to
meson pairs, and the extra $q\qbar$ pairs from the vacuum
form the baryon anti-baryon pair from one of these mesons.
In a phenomenological pole model, they emerge through a meson pole
with a pole mass below threshold, and therefore lead to a
falling matrix element at the threshold.
In a more QCD-motivated picture, the gluons needed to produce these
extra pairs tend to have low $q^2$ due to the running of $\alpha_S$
and the propagator $\propto 1/q^2$.
A typical example is the decay
$B^- \to \Lambda_c^+ \pbar \pi^-$ \cite{Abe2004sr,Aubert2008if}
shown in Fig.~\ref{Fig:barypairmass}a.

A counterexample is the decay
$\Bbar^0 \to \Sigma_c^0 \pbar \pi^+$ shown in
Fig.~\ref{Fig:barypairmass}b where the threshold region is hardly
populated.  Indeed, all diagrams like the example in
Fig.~\ref{Fig:Bbaryfeyn}d would not lead to meson pairs, but rather to
a diquark anti-diquark state.  In a pole model, only a baryon pole could
be assumed to create a baryon meson pair, and the $q\qbar$ pairs
must be created by hard gluons.

While theorists have suggested explanations along these lines
\cite{suzuki}, a rigorous calculation is still missing.

\section*{Acknowledgements}

I wish to thank many colleagues from the BABAR and LHCb collaborations
for helpful discussions on the topics presented here.
I also want to acknowledge this perfectly organized meeting
at \v Strbsk\'e Pleso.


}

\end{document}